\documentclass[twocolumn,prb,showpacs]{revtex4}
\usepackage{psfrag}
\usepackage{graphicx}
\usepackage{amsmath}
\usepackage{amssymb}

\begin{document}

\title{Quantum interference effects in electron transport through nitrobenzene with pyridil anchor groups} 

\author{R.~Stadler}
\affiliation{Department of Physical Chemistry, University of Vienna, Sensengasse 8/7, A-1090 Vienna, Austria}

\date{\today}

\begin{abstract}
We present density functional theory (DFT) based non-equilibrium Green's function (NEGF) calculations for the conductance through a nitrobenzene molecule, which is anchored by pyridil-groups to Au electrodes. This work is building up on earlier theoretical studies where quantum interference effects (QIE) have been identified both in qualitative tight binding and in DFT descriptions for the same molecule with different chemical connections to the leads. The novelty in the current contribution is two-fold: i) The pyridil-anchors guarantee for the conductance to be determined by rather narrow peaks situated closely to the Fermi energy which is relevant because it might maximize the impact of quantum interferences on the I/V behaviour. In a scan of eight different junction setups, where the connection sites of aromatic rings, their torsion angle with respect to each other and the surface structure have been varied, QIE was found to dominate the conductance for only one planar geometry. For finite torsion angles between aromatic rings the effect moves to higher energies and would therefore only be accessible for experimental observation in a gated junction. ii) A detailed comparison between simple topological models and DFT results for the investigated systems aims at assessing the usefulness of such models as analysis tools for a better understanding of the physics of QIE and its structure dependence. 
\end{abstract}
\maketitle

\section{Introduction}

Electron transport between nanoscale contacts has become a much debated field in the last few years, due to its possible applicability in molecular electronics and recent progress in its experimental characterisation~\cite{molelect}. From a theoretical point of view, first-principle non-equilibrium Green's function (NEGF) methods are typically implemented~\cite{atk,xue,sanvito,kristian} in combination with density functional theory (DFT) for describing electron transport through single molecule junctions.\\
Recently, interest in employing quantum interference effects (QIE) for the design of single molecule devices has intensified~\cite{baer1}-\cite{baranger}. Such effects occur when electron waves pass through nano junctions phase-coherently and depend crucially on symmetry. The relevance of QIE has been first observed in the context of waveguides for semiconductor nanostructures~\cite{macucci}-\cite{ballistic}, but it has also been recognized early on theoretically~\cite{joachim1} and experimentally~\cite{joachim2,mayor} that electron transport through a benzene molecule connected in a meta-configuration is drastically reduced when compared to equivalent junctions with ortho- or para-connections between the molecule and the leads. This dependence of the conductance on the connection site in benzene has been identified as QIE and has been explained in terms of phase shifts of transmission channels~\cite{joachim1,tada,ratner}, which are intimately linked to the nodal structure of the involved molecular orbitals (MOs), and it has also been generalized to electron transport through larger aromatic molecules~\cite{baer1,stafford1,stafford2,baranger,baer2}. Interestingly, a similar connection site dependence for benzene has been theoretically proposed for the non phase-coherent Coulomb blockade regime~\cite{coulomb1,coulomb2}.\\
For making use of QIE and its influence on electron transport properties as an enabling tool for single molecule devices two conceptually different proposals have been brought forward: i) In a QIE transistor tunable coherent current suppression can be achieved by applying a local electric field or gate potential to an aromatic molecule covalently bonded to source and drain electrodes, where the operating principle is introducing decoherence or dephasing to the electron transmission between the leads~\cite{baer1,stafford1,stafford2,baranger}. ii) Another way of taking advantage of QIE for molecular electronics lies in controlling electron transmission by chemically modifying or changing the conformation of side groups to aromatic molecules, which has been suggested as a concept for data storage~\cite{first}, the implementation of intra-molecular logic gates~\cite{second} or single-molecule sensors~\cite{lambert}. Due to the typical shape, which interference patterns caused by side groups adopt in the transmission functions, they have by some authors been classified as Fano resonances~\cite{lambert,ernzerhof1,ernzerhof2} in order to place them into the wider context of antiresonances in molecular wires~\cite{emberly}. Most recently, the effect of side groups has also been demonstrated for so-called cross-conjugated molecules~\cite{ratner1}, which appear to be very good candidates for implementing switching and also rectifying behaviour in single molecule junctions, where field gating can be  combined with side group induced interference for the design of stable devices with good operating characteristics~\cite{ratner2}.\\
\begin{figure*}
\includegraphics[width=0.8\linewidth,angle=0]{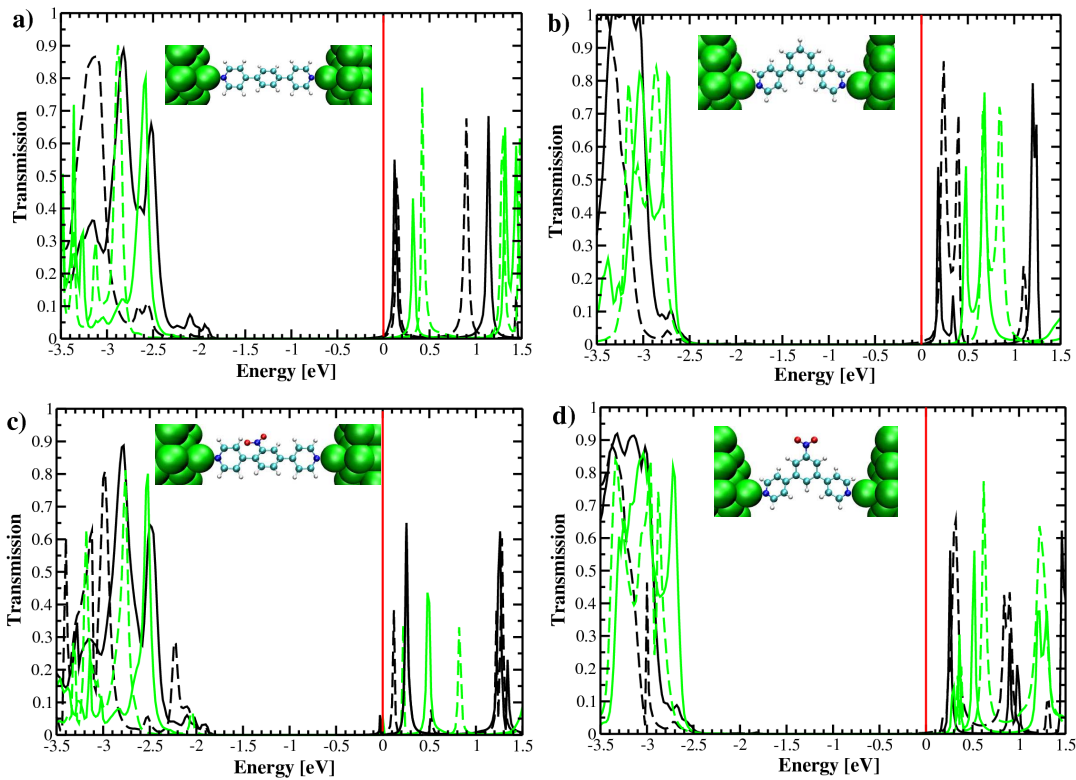}
\caption{\label{fig1}NEGF-DFT transmission functions for a),b) benzene and c,d) nitro-benzene anchored to gold leads via pyridil-groups in a),c) para- and b),d) meta-configurations, respectively. The green/bright lines correspond to calculations with the molecules attached to flat Au fcc (111) surfaces, whereas for the black/dark lines they are bonded to Au-adatoms on top of Au fcc (111). The detailed structures for the latter are shown as insets. The solid and dashed lines distinguish between molecules with planar structures and with torsion angles of a) 37.0$^{\circ}$, b) 33.5$^{\circ}$, c) 41.9$^{\circ}$ and d) 33.8$^{\circ}$ between the aromatic rings, respectively. E$_F$ is the energetic point of reference, i.e. the zero-point on the x-axis.}
\end{figure*}
In our article we focus on the side group proposal for QIE based single molecule devices from our earlier work~\cite{third}, where the influence of nitro-groups on the conductance of benzene directly connected to gold electrodes by thiol groups or separated by poly-acetylene spacers has been investigated. There destructive interferences in the transmission function due to the chemical substitution with NO$_2$ have been encountered as dips in rather broad peaks at $\sim$ 1 eV above the Fermi level E$_F$. This rather large energetic distance of QIE from E$_F$ presents an obstacle for the chosen system to be applicable directly for single molecule devices, and the broadness of the transmission function in the region of interest complicated the analysis. Recent theoretical calculations on bipyridine between gold leads on the other hand, revealed for this more weakly bonded molecule that the conductance is determined by rather narrow lowest unoccupied moleculer orbital (LUMO) peaks situated rather closely to E$_F$~\cite{bipy} in the transmission functions, where rather good agreement with a reliable set of conductance measurements could be achieved~\cite{tao}. This good agreement somewhat diminishes the concerns one might have about the applicability of DFT for weakly coupled systems due to its well-known deficiencies such as self-interaction effects~\cite{perdew,sanvito1} or the lack of a derivative discontinuity in the exchange correlation functional~\cite{perdew1}. One has to keep in mind that the term {\it weak coupling} is a relative one and that it is not straightforward to estimate the impact of common approximations on the quality of predictions in a coupling regime of intermediate strength. The differences in the peak positions between bipyridine and biphenyl-dithiolate could be explained in terms of equilibrium charge transfer~\cite{fermi1}, where the exact values for bipyridine also dependended on the surface structure of the Au leads~\cite{fermi2}. \\
Combining the knowledge gained from Ref.~\cite{third} on the one side and Refs.~\cite{bipy,fermi1,fermi2} on the other side we present in our current contribution NEGF-DFT calculations for the conductance of nitro-benzene connected to Au electrodes by pyridil anchors, where we also focus on the impact of varying configurational degrees of freedom of the molecule and the surface structure. Another goal of this work is to provide a direct semi-quantitative comparison of our DFT results with the predictions of simple topological H\"{u}ckel or tight-binding (TB) models in order to explain the similarities and differences of results found on those two levels of theory and to assess the usefulness of TB as an analysis tool. \\
The paper is organized as follows: In the next section we introduce the computational details of our NEGF-DFT approach and apply this method for calculating the transmission functions for the different molecular and surface structures. In the following section we present topological models mimicking the molecules in this article and provide a semi-quantitative comparison with NEGF-DFT. This comparison we conduct in two steps where the first step is based on the simplistic AO models in Refs.~\cite{first} and ~\cite{second} (NEGF-TB1) with all onsite energies and coupling elements set to 0 and 1, respectively, and side-groups represented by just a single orbital. In a second step we take all atomic orbitals constituting the $\pi$ states of the molecule from DFT calculations in order to design a more realistic Hamiltonian (NEGF-TB2). In Sec.~\ref{sec:cond} we discuss the results obtained for the conductance and explicitly prove the occurence of QIE for one of the investigated junction geometries. For our arguments at this point we make use of another tight binding description which is based on molecular orbitals (NEGF-2MO-TB). In this context we also address the issue of the dependence of the observability of QIE in the chosen systems on the molecular conformation. In the final section a summary and outlook are given.

\section{Structures and technique for NEGF-DFT transmission functions}\label{sec:comp}

\subsection{Molecular and surface structures}

In Fig.~\ref{fig1} we show transmission functions for the variety of structures investigated in this article. The main effects we want to study are destructive interferences close to E$_F$ brought about by a nitro-group when attached to a benzene molecule~\cite{second,third} and their dependence on the connection to the Au leads. For this end the structure of the junction is varied in four ways: i) A para-bipyridil phenyl (para-bpph)~\cite{bpph1,bpph2} (see the inset of Fig.~\ref{fig1}a) is contrasted with a molecule where the two pyridil anchors are connected to the central benzene ring in a meta-configuration (meta-bpph) (Fig.~\ref{fig1}b). ii) We add nitro-groups to para- (Fig.~\ref{fig1}c) and meta-bpph (Fig.~\ref{fig1}d). iii) In their equilbrium structure bpph-molecules exhibit a finite torsion angle, which we have optimized with DFT-total energy calculations for the free molecules. Since the highest occupied molecular orbital (HOMO)-LUMO gaps and conductances of the molecules depend on this angle, we compare transmission functions for planar molecules (solid lines) and their counterparts with realistic torsion angles (dashed lines) in Fig.~\ref{fig1}. The particular case of nitro-para-bpph is exceptional in the sense that there is a substantial total energy difference of $\sim$ 15 eV distinguishing the unstable planar from the geometry optimized finite angle structure due to the close proximity of one oxygen atom of NO$_2$ and one hydrogen atom of a neighbouring pyridil ring within the plane, which leads to rather large repulsive forces in the planar molecule. Therefore, the dependence of our results on the molecular conformation is explicitly investigated in Sec.~\ref{sec:cond} C. For all other molecules in our study the total energy differences between planar and optimized geometries are less than 0.5 eV, meaning the angle could in principle be adjusted by means of chemical synthesis~\cite{torexp}. iv) From our studies of bipyridine~\cite{bipy,fermi1,fermi2} we know that the position of the LUMO with respect to E$_F$ can depend on the surface structure of the leads. We therefore compare a flat Au fcc (111) surface (green/bright lines) with a surface that has an additional Au-ad-atom (black/dark lines), where Au-N distances of 2.42 \AA \ and 2.12 \AA \ have been chosen, respectively, in agreement with earlier findings~\cite{bipy} and the adsorption configuration was assumed to be on-top for all structures. Whereas the geometries of the isolated molecules have been relaxed by total energy minimisation, the Au atoms have been fixed to their positions in the bulk crystal structure throughout our study. Although we therefore neglect the effect that the electrode-molecule interaction has on the structure of both molecule and surfaces, we believe that the impact of this effect on the transmission function should be minor, when compared to the configurational degrees of freedom we explore in this article.\\

\begin{figure*}
\includegraphics[width=0.8\linewidth,angle=0]{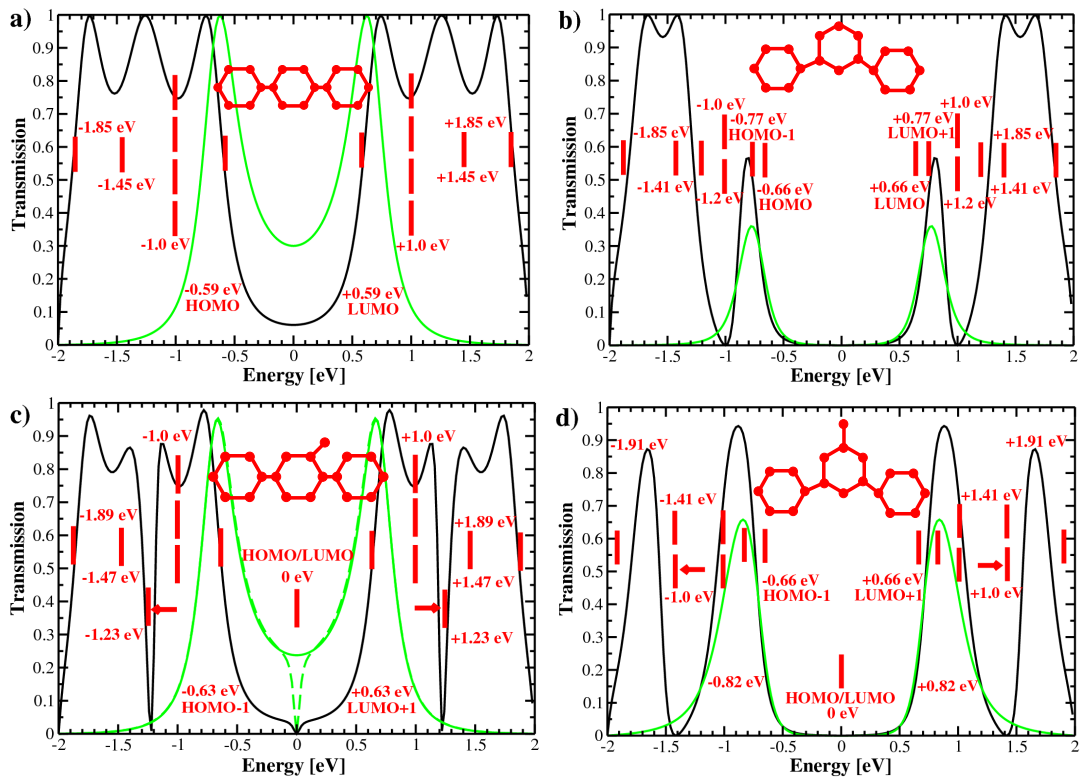}
\caption{\label{fig2}NEGF-TB1 transmission functions mimicking the topology of $\pi$ electrons of a),b) bpph and c,d) nitro-bpph anchored to semi-infinite chains with all onsite energies H$_{ii}$= 0 eV and coupling matrix elements H$_{ij}$= -1 eV for nearest-neighbour atoms connected by a chemical bond. The black solid lines show the total transmission, the green solid lines only the contributions coming from a) HOMO and LUMO, b) HOMO-1, HOMO, LUMO and LUMO+1, c) HOMO-1 and LUMO+1 and d) HOMO-2, HOMO-1, LUMO+1 and LUMO+2, respectively. For the green dashed lines in c),d) the additional state due to the orbital directly at the Fermi energy is also included. As insets schemes of the molecules and the energetic distributions of molecular orbitals are also displayed. E$_F$ is the energetic point of reference, i.e. the zero-point on the x-axis.}
\end{figure*}

\subsection{Computational details}

For all DFT calculations in this work we have used the Siesta code~\cite{siesta} with a double-zeta polarized (DZP) basis set, Troullier-Martins type norm-conserving pseudopotentials~\cite{martins} for all elements and a Perdew-Burke-Ernzerhof (PBE) parametrization~\cite{perdew2} for the exchange correlation (XC) functional. Our NEGF-DFT approach for the electron transport problem has been described in detail in Refs.~\cite{kristian} and~\cite{mikkel}. It requires a division of the single molecule junctions into three regions, namely a left lead, a right lead and a scattering region. For all three regions independent supercell calculations with periodic boundary conditions are performed, where in the plane perpendicular to the transport direction 3x3 sections of Au fcc (111) form the unit cell. For the molecules studied in this article this means that some of them are only a few \AA \ apart in distance within the surface plane, which makes the setup more akin to self-assembled monolayers sandwhiched between two electrodes rather than a single-molecule junction. This distinction is significant because the amount of coverage can have an impact on the alignment of molecular levels with the Fermi energy due to dipolar interaction between molecules or screening effects~\cite{romaner}. We expect, however, that the dependence of the occurence or absence of QIE on the molecular structure will be unaffected by such considerations. A total of seven gold layers have been incorporated into the scattering region in oder to insure a convergence of the effective potential and electron density to bulk values. The leads are then added to the Green's function of the scattering region as self energies which allows for the calculation of transmission functions using the Keldysh formalism~\cite{keldysh}. We found that {\bf k}-point sampling in the transverse plane is important in order to obtain welll converged results for the conductance~\cite{kpoints} and used a 4x4 grid for all systems but checked for one structure that 6x6 and 8x8 grids did not change our results in any way. \\

\subsection{Structure dependence of transmission functions on a large energy scale}

In relation to Fig.~\ref{fig1} we now discuss overall features of the transmission functions with an emphasis on the structural variations i), iii) and iv) as quoted above. A detailed analysis of the conductance and interference effects will be delivered in upcoming sections of this article. Here we want to point out that the structural differences, which are marked by line-types, namely the surface geometry (bright vs. dark) and the torsion angle (solid vs. dashed) are very similar for all panels in Fig.~\ref{fig1}. A planar molecule means a maximum of $\pi$ conjugation across the molecule, which results in a minimum of the HOMO-LUMO gap. Introducing a torsion angle of 33-42$^{\circ}$ opens up this gap, which can be seen for all molecules when moving from the solid to dashed lines and which is in agreement with earlier studies on bipyridine~\cite{bipy}. The Au-ad-atom on the Au (111) fcc surface on the other side has the effect of moving the LUMO closer to E$_F$ in energy as has been explained in Ref.~\cite{fermi2} and which is again not influenced by our variations of the structure of the molecule. Finally, we observe that molecules connected in a para-configuration (Figs.~\ref{fig1}a and c) exhibit smaller HOMO-LUMO gaps and a LUMO closer to E$_F$ than the meta-connected junctions (Figs.~\ref{fig1}b and d).

\section{Tight binding predictions and their relation to NEGF-DFT results}

The NEGF formalism for electron transport can also be combined with a topological H\"{u}ckel model for the conjugated $\pi$ electrons in aromatic systems, where the Hamiltonian matrix is defined by single atomic orbitals (AOs) for each site in the molecule, which is occupied by a carbon atom, and with finite coupling elements to their respective nearest neighbours only. Such a simplified description has often been used for calculating transmission functions of single molecules (where the leads are also approximated by chains of AOs within NEGF) when design schemes for electronic devices have been proposed~\cite{baer1,first,second,stafford1,stafford2}, because it allows for an analytical derivation of general principles~\cite{first,second}, numerical calculations for large systems and a straightforward phenomenological inclusion of many-body effects~\cite{kristian1}. The latter are beyond the scope of this article, which is focused on effects in the coherent transport regime. 

Calculations in a topological H\"{u}ckel framework can always only be a first step or proof-of-principle, where for realistic quantitative predictions, which can then be compared with related experiments more accurate NEGF-DFT results are needed taking all the details of the atomic and electronic structure of the molecule and leads into account~\cite{third,baranger}. It is, however, instructive to see the qualitative and quantitative correspondence or, respectively, the sources of discrepancies between NEGF-DFT and simple models for a given range of systems. In this section we introduce calculations based on the H\"{u}ckel model on two levels of approximations: i) For NEGF-TB1 all onsite energies are 0 eV (meaning that all AOs are at E$_F$) and all nearest-neighbour couplings are set to -1 eV. As a further approximation we make no difference between carbon, nitrogen and gold sites and neglect the presence of oxygen and hydrogen. In this way no information is taken from DFT calculations, experiments or chemical data in the literature. Hence the Hamiltonian is parameter free in the sense that all parameters are set to equal (and arbitrary) values and such a simple theory can provide the information how far the molecular topology alone can affect the occurence of QIE when all more subtle details of the electronic structure are disregarded. In a second step one can also ask the question what are the key parameter changes required for enhancing the qualitative and quantitative agreement between NEGF-DFT and NEGF-TB1. We will demonstrate this for para-bpph below. ii) For NEGF-TB2 we make use of the availability of a set of realistic parameters for onsite energies and couplings from an analysis of the DFT calculations. Here we go a step further and ask the question how far the more basic assumptions of the toplogical H\"{u}ckel model, namely the description of all atoms (including the leads) by single AOs with only nearest-neighbour coupling and a disregard of even the atomic structure of the leads, affect the transmission function qualitatively and quantitatively, when chemically reasonable parameters have been obtained for the model Hamiltonian.\\

\subsection{Topological mimicking of molecular structure as the simplest level of approximation -NEGF-TB1}

\begin{figure}
\includegraphics[width=0.8\linewidth,angle=0]{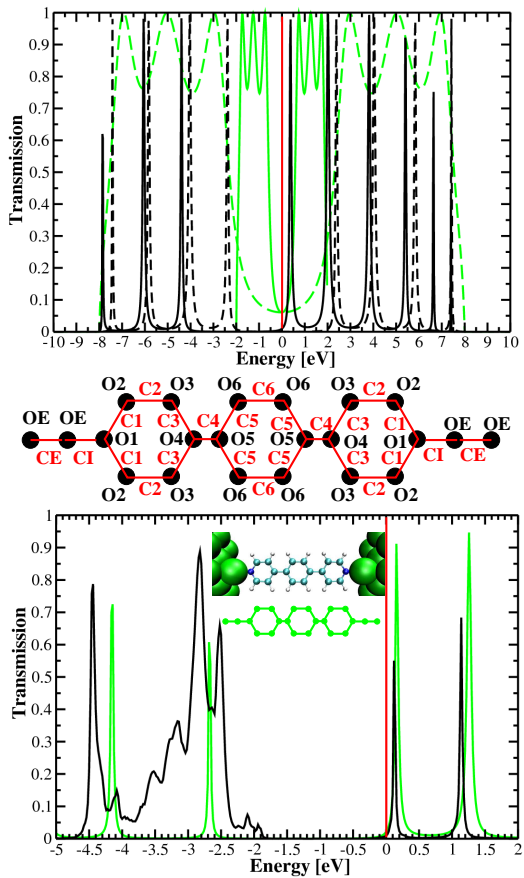}
\caption{\label{fig3} (Top) Step-wise variation of the H\"{u}ckel parameters of unsubstituted para-bpph (Fig.~\ref{fig2}a) in NEGF-TB1 for getting closer to the NEGF-DFT results for the transmission function. Green solid line: all H$_{ii}$= 0 eV and all H$_{ij}$= -1 eV, green dashed line:  as before but all H$_{ij}$= -4 eV, black dashed line: as before but CI = -1 eV, black solid line: as before but O1-O6 = -2 eV. (Middle) short-hand notation for non-symmetry equivalent onsite energies H$_{ii}$ and coupling elements H$_{ij}$ for para-bpph. (Bottom) Direct comparison of full NEGF-DFT results and NEGF-TB2 calculations, where the parameters for the molecular Hamiltonian have been derived directly from Siesta~\cite{siesta} orbitals of C-p-states perpendicular to the aromatic plane and only the leads are approximated with OE = 0 eV, CE = -10 eV and CI = -1 eV.}
\end{figure}

For the side-group approach to QIE based devices, rather strong interference effects have been found directly at E$_F$ within NEGF-TB1~\cite{first,second} and a more complicated picture emerged when using NEGF-DFT~\cite{third}. In Fig.~\ref{fig2} we present NEGF-TB1 calculations for one-orbital systems mimicking the molecules in Fig.~\ref{fig1}. By subdiagonalizing this Hamiltonian for the ``molecular" AOs only~\cite{kristian,third}, we can derive the energy eigenvalues of molecular orbitals (MOs), which have been placed as vertical lines with respect to the energy axis in Fig.~\ref{fig2}. Now we can also decouple MOs from the leads and the green/bright lines show the contribution to the transmission functions, which comes from the MOs closest to E$_F$ only (see the caption of Fig.~\ref{fig2} for details), which for the interesting energy range qualitatively reproduces the dependence of the total transmission (black/dark lines) on molecular topology. This means that for para-bpph the conductance, i.e the transmission at E$_F$ is finite (Fig.~\ref{fig2}a), and is reduced to zero by destructive interference when a side-orbital is added (Fig.~\ref{fig2}c). For meta-bpph on the other side the conductance is zero already in the unsubstituted molecule (Fig.~\ref{fig2}b) and adding a side-orbital therefore makes no difference (Fig.~\ref{fig2}d).\\
We note that for the NEGF-DFT transmission functions (Fig.~\ref{fig1}) we find molecular HOMO-LUMO gaps for all systems and the conductance is determined by the fine structure of the LUMO-peak and its distance from E$_F$, which contrasts with the simplistic picture derived from Fig.~\ref{fig2}. There are three discrepancies in particular: i) The band-width of the transmission function in NEGF-TB1 is too small compared with NEGF-DFT; ii) there is just one broad peak-structure in NEGF-TB1 instead of the narrow and isolated peaks found above the Fermi level in NEGF-DFT which start from a distinct HOMO-LUMO gap; and iii) in DFT E$_F$ is close to the LUMO peak, whereas it is in the middle of the gap in NEGF-TB1. It is therefore instructive to interpolate between the two, which is illustrated for unsubstituted para-bpph in Fig.~\ref{fig3}. In the following we adopt the short-hand notation for onsite energies (OE, O1-O6) and coupling elements (CE, CI, C1-C6) outlined in the diagram in the middle of Fig.~\ref{fig3}. As can be seen in the top graph of the figure in removing the discrepancy i) the transmission function covers a larger energy range if all coupling elements are changed from -1 eV to -4 eV (moving from the solid green to the dashed green line). This is because the band dispersion of the single semi-infinite state formed by the AOs representing the leads is proportional to the coupling between them. When now only the coupling between the molecules and the leads CI is set back to -1 eV (black dashed line) a regular structure of narrow MO peaks with a HOMO-LUMO gap appears, which means getting rid of discrepancy ii) from the list above. Finally, by moving the onsite energies O1-O6 from 0 eV to -2 eV (which is more realistic for the position of C-2p-orbitals with respect to the Fermi level of the Au electrodes), the LUMO gets very close to E$_F$ (black solid line) and thereby also the source for discrepancy iii) has been identified. 

\subsection{A hybrid model combining TB simplicity with DFT parameters - NEGF-TB2}

In what we call NEGF-TB2 in this article the H\"{u}ckel description of the molecule is derived from NEGF-DFT calculations directly~\cite{kristian,mikkel}, where the Hamiltonian and overlap matrices are defined by the sub-space spanned by the C-2p-orbitals perpendicular to the molecular plane in the basis set of Siesta calculations~\cite{siesta}. The bottom part of Fig.~\ref{fig3} compares the results from NEGF-TB2 and NEGF-DFT for para-bpph in a planar configuration. The peaks of NEGF-TB2 (solid green lines) coincide with the centers of the peak structure of the NEGF-DFT transmission function. The deviations in peak shapes can readily be explained by the differences in the description of the leads, where the broad structure with its center at $\sim$ -2.7 eV can be attributed to the Au-d-states, which are lacking in the AO-chains of NEGF-TB2. Above E$_F$, however, the Au-s-states are dominant and the two calculations agree very closely even in peak widths. This is even more astonishing since for the NEGF-DFT curve all MOs are contributing to the transmission, whereas for NEGF-TB2 only the states built from C-2p-AOs perpendicular to the molecular plane are considered.

\subsection{Discussion}

Although the topological H\"{u}ckel model in its most simplified form (Fig.~\ref{fig2}) results in transmission functions that differ significantly from DFT data (Fig.~\ref{fig1}), it is possible to reconcile the model with NEGF-DFT by the choice of realistic parameters for onsite energies and couplings as achieved in NEGF-TB2 (Fig.~\ref{fig3}). Such parameters can only be obtained from the full DFT-calculation, since their values are highly sensitive on various aspects of the junctions atomic and electronic structure. Therefore, the usefulness of calculations based on a topological H\"{u}ckel model lies not so much in being computationally less demanding than NEGF-DFT but rather in its ability of reducing the complexity of the system as a tool for analysis, where the number of orbitals that have a distinctive effect on the transmission function can be reduced to a minimum, in the shown case just one C-2p-AO per carbon atom. This might in principle allow for an optimization of desired effects such as QIE, by taking the onsite energies and couplings from DFT-calculations for a given molecule, varying them in a systematic way and searching for a different molecule afterwards, which embodies the changed pattern of parameters. In praxis, however, this would be a formidable task, because a change in molecular structure is likely to vary not only one or a few TB parameters but most of them at the same time.\\
Nevertheless, there might be another benefit of simplified topological models without realistic parametrisation in terms of qualitative rather than quantitative predictions. We will show in the next section that, although the transmission functions derived from NEGF-TB1 differ widely from the ones obtained from NEGF-DFT, the basic topology related result of Fig.~\ref{fig2}, namely that QIE is found for nitro-bpph in a para-connection (Fig.~\ref{fig2}c) but not for the meta-geometry (Fig.~\ref{fig2}d), also applies for the ab initio values for the conductance. This indicates that the occurence or absence of QIE is a purely topological feature of the molecular structure, while the finer details of the atomic and electronic structure of the junction on the other hand determine at which energies such effects can be found in the transmission function and whether they are observable by experiments.

\section{Interference effects and their impact on the conductance}\label{sec:cond}

\subsection{Structure dependence of the zero bias conductance}

\begin{figure*}
\includegraphics[width=0.8\linewidth,angle=0]{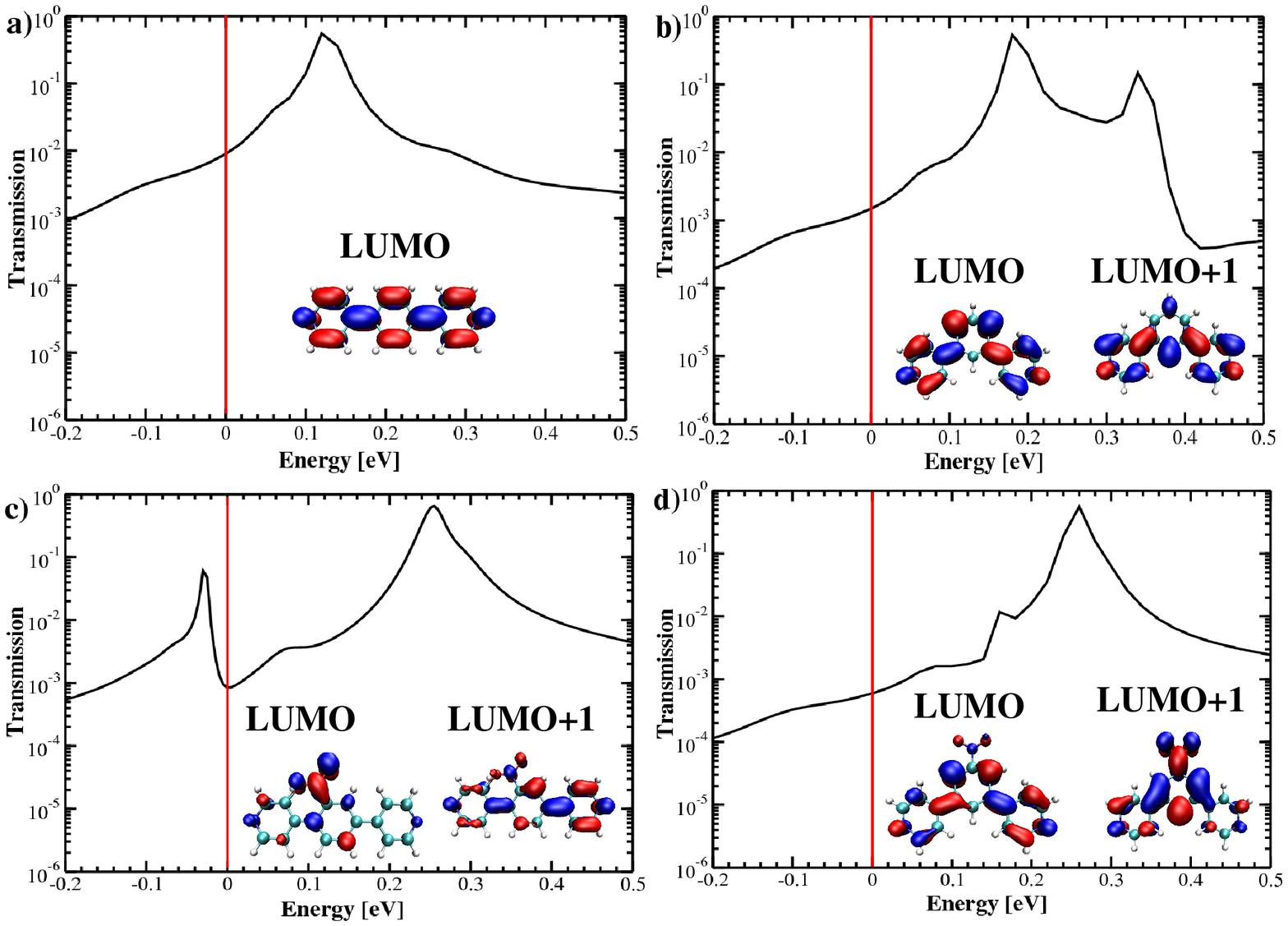}
\caption{\label{fig4}Logarithmic plots of NEGF-DFT transmission functions for the same structures and with the same notations as in Fig.~\ref{fig1}, but here we zoom into a much narrower energy range around E$_F$ and show only results for planar molecules attached to surfaces with Au-adatoms. Spatial distributions of relevant MOs are shown as insets.}
\end{figure*}
\begin{table*}
\caption{\label{cond.tab}Numerical values for the conductance taken from the transmission functions at E$_F$ for a surface structure with Au ad-atoms and planar molecules (corresponding to Fig.~\ref{fig4}). The predictions of the NEGF-TB1 model in Fig.~\ref{fig2} are also given for comparison. These predictions are only reflected by the NEGF-DFT data if the aromatic molecules are planar and therefore the conjugation of $\pi$ electrons is maximal as outlined in the text.}
\begin{tabular}{c|cccl} \hline
 & \multicolumn{4}{c}{Conductance in multiples of 10$^{-3}$ G$_0$} \\
& para-bpph  & nitro-para-bpph & meta-bpph & nitro-meta-bpph \\
\hline
NEGF-DFT & 9.5 & 0.8 & 1.5 & 0.6 \\
qualitative TB predictions & ON & OFF & OFF & OFF \\
\hline
\end{tabular}
\end{table*}

We now turn our attention to the zero-bias conductance of the structures introduced in Fig.~\ref{fig1}, which is defined as the numerical value of the transmission function at E$_F$. For the calculations with flat surfaces as leads the conductance is in general an order of magnitude smaller for molecules bound by pyridil-linkers due to the larger energetic distance of the LUMO to E$_F$~\cite{fermi2} and therefore less suitable for accurate predictions within our approach, where the boundary conditions for the exponential decay of basis functions can play a significant role for small values~\cite{mikkel}. The effect of finite torsion angles between aromatic rings, which tend to disrupt the conjugation of the $\pi$ system and thereby open up the HOMO-LUMO gap will be investigated in more detail in Sec.~\ref{sec:cond} C. In general it was found previously that for such tilted systems contributions from $\sigma$ electrons gain in relevance for the transmission~\cite{ratner3}.\\ 
Fig.~\ref{fig4} is based on the same calculations as Fig.~\ref{fig1} but displays a more narrow energy range on a logarithmic scale thereby allowing to discern the peak structure mostly caused by the LUMO and LUMO+1. Also here it can be seen that due to the larger gap-size for meta-bpph with and without nitro-substitution the molecular peaks are somewhat higher above the Fermi level than for para-bpph. This does not necessarily determine the conductance alone as will be shown later on, since the latter quantity is also defined by the broadness of the transmission peaks and the magnitude of their tails at E$_F$.\\
A further distinction between para- and meta-bpph can be found in the effect that the NO$_2$ substitution has on the transmission. For the unsubstituted para-bpph the energy range around E$_F$ is dominated by a single peak which can be attributed to the LUMO (Fig.~\ref{fig4}a), quite similarly to what has been observed for bipyridine~\cite{bipy,fermi1,fermi2}. The introduction of the nitro-group shifts this state up to higher energies and inserts another -- due to weak coupling rather small -- peak mostly localized on NO$_2$ directly at E$_F$ (Fig.~\ref{fig4}c). This peak (although related to the LUMO) appears even slightly below E$_F$ for the planar molecules as a consequence of the hybridisation between molecular and surface states. It is interesting to note that only this peak exhibits a lineshape which is comparable to the Fano resonances identified in Ref.~\cite{lambert}. For meta-bpph on the other hand, the transmission spectrum directly at E$_F$ is governed by the interplay of two MOs with a spatial localisation pattern which does not change considerably when moving from the unsubstituted (Fig.~\ref{fig4}b) to the nitro-substituted (Fig.~\ref{fig4}d) molecule, although there are some differences in the shape of the transmission function in the sense that the two distinct peaks move energetically closer to each other in Fig.~\ref{fig4}d. 
In Table \ref{cond.tab} we list the calculated conductance values for surface structures with Au-ad-atoms, where at least for planar molecules with maximal conjugation of $\pi$ electrons, the results from NEGF-DFT reflect the NEGF-TB1 predictions. With respect to the predictive power of our results, we have to point out that neither the torsion angle between aromatic rings nor details of the surface structure, both having a crucial impact in our data, can be easily controlled in actual experiments. The aims of our current study are rather to elucidate the basic physics of QIE and their structure dependence and we do not claim at this stage to have arrived at the design of molecules for robust devices.\\ 

\begin{figure}
\includegraphics[width=0.8\linewidth,angle=0]{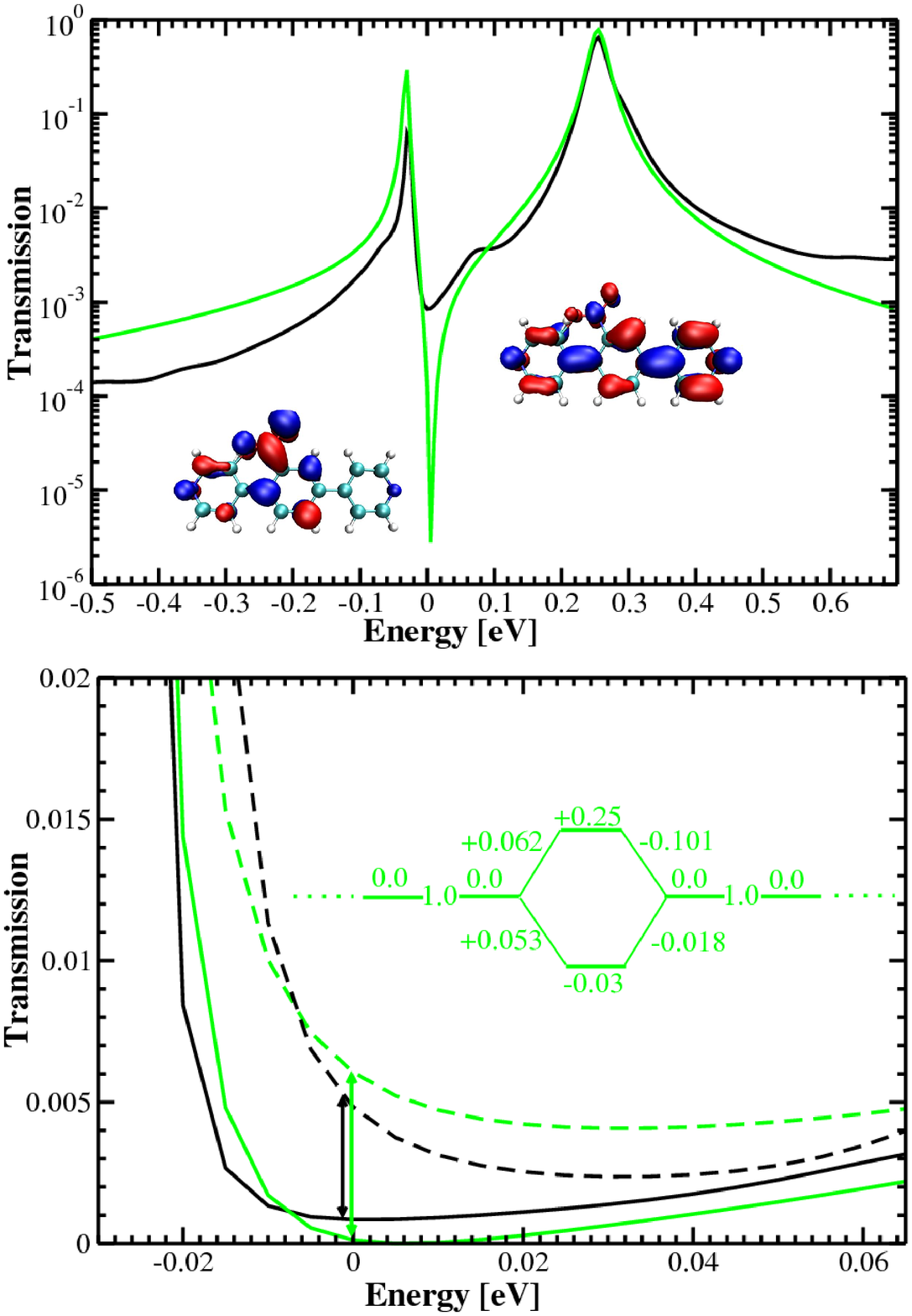}
\caption{\label{fig5}Transmission functions for nitro-para-bpph in a planar configuration (corresponding to the black dotted line in Fig.~\ref{fig4}). (Top) The full NEGF-DFT calculation (solid black line) is shown to be reproduced by a NEGF-2MO-TB model (green solid line, see text for details) in a logarithmic plot. The insets show the spatial distributions of the two MOs in question. (Bottom) Interference effects are quantified by comparing the full transmission (solid lines) with the sum of the transmissions through the two MOs (dotted lines) for the full NEGF-DFT (black) and the NEGF-2MO-TB-model (green). The inset defines the setup for the NEGF-2MO-TB-model.}
\end{figure}

\subsection{Interference effects in the NEGF-DFT results and their analysis with a NEGF-2MO-TB-model}

In order to demonstrate that the reduction of the conductance of para-bpph by the substitution with NO$_2$ is indeed due to interference effects, we follow Ref.~\cite{sautet} in defining an effective coupling $\Gamma$ 
\begin{equation}
\Gamma=\frac{\alpha_{A}\beta_{A}}{E - \epsilon_{A}} + \frac{\alpha_{B}\beta_{B}}{E - \epsilon_{B}}
\end{equation}
for describing the transmission at a given energy E (which would normally be the Fermi energy E$_F$ for an ungated system if the focus is on the zero-bias conductance) as the cumulative effect of two energetically nearby MOs with the indices A and B, where $\epsilon_{A,B}$ denote their eigenenergies in the environment of the junction, and $\alpha_{A,B}$ and $\beta_{A,B}$ their couplings to the left and right leads, respectively. In Fig.~\ref{fig5} we show that such a simple 2MO-model when combined with a TB-description of the leads within the NEGF-formalism captures all the characteristics of the transmission peaks calculated closely above E$_F$ for planar nitro-para-bpph from NEGF-DFT and give explicit values for $\epsilon_{A,B}$, $\alpha_{A,B}$ and $\beta_{A,B}$. In the upper part of the figure (shown on a logarithmic scale) we demonstrate that the distinct Fano-type lineshape~\cite{lambert} of the NEGF-2MO-TB-model (bright solid line) can also be found although apparently slightly diminished in the NEGF-DFT results (black solid line). The differences between the two curves can be readily understood when it is considered that the results from the model describe electron transport through the two MO's in question only (hence the bright line goes down to zero at the minimum), whereas for the NEGF-DFT calculations orbitals energetically further apart give a negligible but on a logarithmic scale visible contribution to the transmission~\cite{bipy}. From the lower part of Fig.~\ref{fig5} it can be seen that when we add up the transmission through MOs A and B in isolation from each other, that is by decoupling them in turn from the leads, we arrive at a conductance that is considerably higher than the one with both orbitals contributing at the same time. This behaviour is found for both, the NEGF-2MO-TB-model and the NEGF-DFT simulations, and an explanation can be instantly found from the definition of $\Gamma$ above. Since for the addition of the terms coming from MOs A and B, the values for the couplings $\alpha_{A,B}$ and $\beta_{A,B}$ have the same sign in both cases, the difference in sign for $\epsilon_{A}$ and $\epsilon_{B}$ stemming from their energetic location being on different sides of E$_F$, results in an effective cancellation of the two contributions. This is the hallmark of destructive interference in the transmission of electron waves.

\subsection{Conformation dependence of QIE in nitro-para-bpph}

\begin{figure}
\includegraphics[width=0.8\linewidth,angle=0]{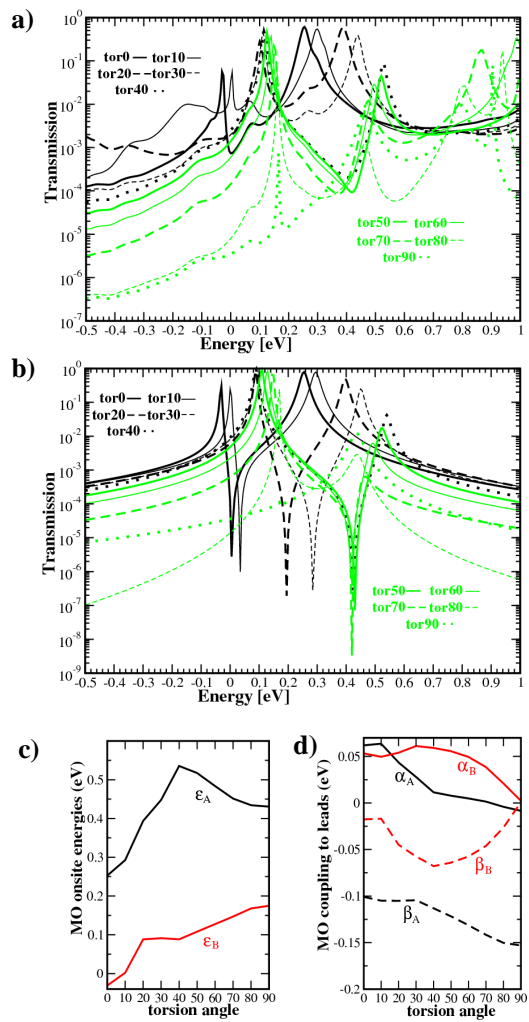}
\caption{\label{fig6}
Transmission functions for nitro-para-bpph in dependence of the torsion angle between the inner benzene and the two pyridil rings calculated from a) NEGF-DFT and b) NEGF-2MO-TB (as introduced in Fig.~\ref{fig5}). The colour code for the line type in dependence on the angles is the following: 0$^\circ$ - thick black solid, 10$^\circ$ - thin black solid, 20$^\circ$ - thick black dashed, 30$^\circ$ - thin black dashed, 40$^\circ$ - thick black dotted, 50$^\circ$ - thick green solid, 60$^\circ$ - thin green solid, 70$^\circ$ - thick green dashed, 80$^\circ$ - thin green dashed and 90$^\circ$ - thick green dotted. The torsion dependence of the onsite energies ($\epsilon_A$ and $\epsilon_B$) and coupling elements ($\alpha_A$,$\beta_A$,$\alpha_B$ and $\beta_B$) as the key parameters for NEGF-2MO-TB is shown explicitly in c) and d), respectively.}
\end{figure}

As explained in Sec.~\ref{sec:comp} the planar molecular conformation of nitro-para-bpph is energetically highly unstable due to direct steric repulsion between one of the oxygen atoms of NO$_2$ and one of the hydrogen atoms of the pyridil ring next to it. In principle the torsion angle can be tuned for systems of neighbouring aromatic rings by introducing electronically inactive aliphatic linker groups as has been demonstrated by recent experiments~\cite{torexp}. This remedy does, however, only allow to overcome fairly small energy barriers, which for the special case of nitro-para-bpph means that such chemical tuning would be only possible in the range of torsion angles of 30$^\circ$ to 90$^\circ$. Therefore, the question arises what happens to QI as predicted in this article in dependence on torsion and what this means for the experimental observability of the effect.\\
In Figs.~\ref{fig6}a and b the transmission functions are shown in dependence on the molecular conformation as calculated from NEGF-DFT and NEGF-2MO-TB, respectively. In Figs.~\ref{fig6}c and d the evolution of the key parameters in NEGF-2MO-TB with the torsion angle is displayed. When comparing the two sets of results two main differences can be identified: i) While in NEGF-2MO-TB (Fig.~\ref{fig6}b) the Fano-type resonance can be distinctly seen for all angles, it appears diminished for low angles in general and disappears completely for 10$^\circ$, 20$^\circ$ and 30$^\circ$ in NEGF-DFT (Fig.~\ref{fig6}a). This discrepancy is likely to be due to contributions coming from $\sigma$ bond channels which compete with the $\pi$ electron mediated electron transport in the energy range of very low transmission~\cite{ratner3}. We note, however, that also for the more accurate NEGF-DFT data, the resonance due to QIE remains clearly observable for the synthetically achievable range of angles from 40$^\circ$ to 70$^\circ$. ii) At the lower and upper borders of the energy range chosen for the figure there is an increase in transmission related to $\pi$ orbitals below the LUMO and above the LUMO+1. These effects cannot be expected to emerge from a 2MO-model, which by definition only describes the contribution of the latter two orbitals, but are without consequence for the subject discussed here.\\
In the relevant features of the conformation dependence, however, Figs.~\ref{fig6}a and b agree rather well. The observed upward shift of the Fano-resonance is due to a shift in onsite energies of the MOs (Fig.~\ref{fig6}c), which must be related to an angle dependence of the charge transfer between the molecule and the electrodes~\cite{fermi1,fermi2}. This upward shift means in practical terms that QIE would be only observable in this molecule experimentally when all levels are shifted by introducing a third electrode as a gate or by electro-chemical gating. There is also agreement between NEGF-DFT and NEGF-2MO-TB on the complete disappearance of QIE at an angle of 90$^\circ$. This is due to the conformation dependence of the couplings between MOs and electrodes~\cite{tortheor}, which apart from $\beta_A$ all become zero in the case of perpendicular aromatic rings (90$^\circ$ in Fig.~\ref{fig6}d). QIE remains observable, however, for torsion angles as high as 70$^\circ$ in Figs.~\ref{fig6}a and b.

\section{Conclusions}

In summary, we presented NEGF-DFT calculations for the electron transport through metal-molecule-metal junctions, where benzene and nitro-benzene have been attached to gold electrodes via pyridil anchor groups. The motivation for this work were proposals for the exploitation of QIE caused by chemical side-groups for the design of active devices in molecular electronics~\cite{first,second}, where at an earlier attempt to implement such a scheme with realistically described chemical systems the broad transmission peak structure of thiol-bonded molecules was found to be an obstacle~\cite{third}. The advantage of the pyridil anchors lies in their weaker coupling to the Au surfaces resulting in narrow peaks in the transmission functions, which are closely above E$_F$ in energy. We could demonstrate that the substitution of the benzene ring in para-bpph with NO$_2$ causes interference effects, which have a high impact on the zero-bias conductance when aromatic rings within the same molecule were in a planar configuration. For conformations with a finite torsion angle the Fano resonance characteristic for QIE in this system moves up to higher energies and a gate would have to be introduced in order to make it observable for experiments. If the molecule is attached to electrodes with a flat surface (i.e. without ad-atoms), the whole peak structure is also shifted upwards in energy, which has been explained in earlier work~\cite{fermi2} in terms of zero-bias charge transfer.\\
In order to faciliate future theoretical design attempts of suitable candidates for device applications, we devoted a substantial part of our study to a comparison of our NEGF-DFT results with topological H\"{u}ckel models. There we found that although parameters derived from full DFT-calculations are needed for a reasonable prediction of the overall shape of the transmission function, also very simplified models predict the occurence of QIE in para-connected nitro-bpph and its absence in the meta-configuration.

\begin{acknowledgements}

The author is currently supported by the Austrian Science Fund FWF, project Nr. P20267. Helpful discussions with Jan Zabloudil, Duncan Mowbray, Mikkel Strange and Victor Geskin are gratefully acknowledged.

\end{acknowledgements}


\bibliographystyle{apsrev}

\end{document}